\documentclass[aip,rsi,reprint]{revtex4-2}

\usepackage{amsmath}
\usepackage{graphicx, amssymb}
\usepackage{xcolor}
\usepackage{epsfig}
\usepackage{xcolor}

\begin{document}

\preprint{APS/123-QED}

\title{Development and Application of a Magnetoelectric Coupling Measurement System for Strain-Mediated Effects}

\author{Samanway Mohanta}
\address{UGC-DAE Consortium for Scientific Research, Indore-452001, India}
\author{Poonacha C.T}
\address{Maharaja Institute of Technology Mysore, Karnataka-570028, India}
\author{Hemant Singh}
\address{UGC-DAE Consortium for Scientific Research, Indore-452001, India}
\author{Gunjan Verma}
\address{UGC-DAE Consortium for Scientific Research, Indore-452001, India}
\author{A. Ahlawat}
\address{Institute of Sciences, SAGE University, Indore-452020, India}
\author{S.C. Das}
\address{UGC-DAE Consortium for Scientific Research, Indore-452001, India}
\author{G. Sinha}
\address{Raja Ramanna Centre for Advanced Technology, Indore-452013, India}
\author{V. G. Sathe}
\address{UGC-DAE Consortium for Scientific Research, Indore-452001, India}
\author{D. K. Shukla}
 \email{dkshukla@csr.res.in}
\address{UGC-DAE Consortium for Scientific Research, Indore-452001, India}

\date{\today}

\begin{abstract}
	This paper reports about development of a dynamic lock-in based magnetoelectric (ME) coupling measurement system and its application to investigate strain-mediated coupling in 0.65Pb(Mg$_{1/3}$Nb$_{2/3}$)O$_3$-0.35PbTiO$_3$/NiFe$_2$O$_4$ (PMN-PT/NFO) and CoFe$_2$O$_4$/BaTiO$_3$ (CFO/BTO) composites. The PMN-PT/NFO composite exhibits a significantly enhanced ME coefficient ($\alpha_{\mathrm{ME}} \approx 1.85$~mV\,mm$^{-1}$\,Oe$^{-1}$) compared to CFO/BTO ($\approx 0.147$~mV\,mm$^{-1}$\,Oe$^{-1}$). The antisymmetric field dependence and $180^\circ$ phase reversal confirm the strain-driven origin of coupling in both the samples. The superior response of PMN-PT/NFO arises from the combined effect of high piezoelectric strength in PMN-PT and low magnetic anisotropy of NFO, whereas the high magnetic anisotropy of CFO and weaker piezoelectricity of BTO limit the ME output. These results highlight the decisive role of magnetic softness and piezoelectric strength in optimizing strain-mediated ME coupling.
\end{abstract}

\maketitle

\section{\label{sec:level1}INTRODUCTION}
Magnetoelectric coupling is a fundamental phenomenon in multiferroic materials in which magnetic and electric order parameters are intrinsically interconnected~\cite{xu2023magnetoelectric,spaldin2005renaissance,nan1994magnetoelectric}. This coupling enables control of magnetization by an electric field and control of polarization by a magnetic field, offering both rich fundamental physics and promising technological applications~\cite{van2004origin,heron2014deterministic}. The possibility of cross-control between magnetic and electric states has stimulated extensive research toward multifunctional devices such as magnetic sensors, actuators, spintronic components, and energy efficient memory systems~\cite{eerenstein2006multiferroic}. A comprehensive understanding of the microscopic mechanisms governing ME coupling is therefore essential for both scientific advancement and device optimization.

The strength of magnetoelectric interaction is quantified by the ME coupling coefficient ($\alpha_{\mathrm{ME}}$), defined as
\begin{equation}
	\alpha_{\mathrm{ME}} = \frac{dE}{dH_{\mathrm{DC}}},
\end{equation}
which represents the change in induced electric field ($E$) in response to an applied DC magnetic field ($H_{\mathrm{DC}}$). Several experimental techniques have been developed to determine $\alpha_{\mathrm{ME}}$, including direct and converse methods, resonance techniques, piezoelectric detection approaches, scanning probe microscopy, and thermal measurements~\cite{vopson2017measurement,hassanpour2022magnetoelectric}. Among these, the dynamic lock-in technique has emerged as a highly sensitive and reliable method, particularly for room temperature measurements and weak ME responses~\cite{duong2007lock,grotel2023application}.

In the dynamic method, a small AC magnetic field is superimposed onto a DC magnetic field, and the resulting electric voltage is detected using a lock-in amplifier (LIA). The induced ME voltage can be expressed as a polynomial expansion in terms of the applied DC magnetic field~\cite{kowal2015magnetoelectric}:
\begin{equation}
	V = \mathrm{Const.} + \alpha H_{\mathrm{DC}} + \beta H_{\mathrm{DC}}^2 + \cdots ,
\end{equation}
where $\alpha$ and $\beta$ correspond to the linear and quadratic ME coefficients, respectively. When a sinusoidal AC magnetic field $H_{\mathrm{AC}}\sin \omega t$ is added to $H_{\mathrm{DC}}$, the total magnetic field becomes
\begin{equation}
	H = H_{\mathrm{DC}} + H_{\mathrm{AC}}\sin \omega t.
\end{equation}
The lock-in amplifier selectively detects the voltage component oscillating at frequency $\omega$, yielding
\begin{equation}
	V_{\mathrm{OUT}} = H_{\mathrm{AC}} \left(\frac{dV}{dH_{\mathrm{DC}}}\right).
\end{equation}
Thus, the magnetoelectric coefficient can be determined as
\begin{equation}
	\alpha_{\mathrm{ME}} = \frac{1}{d}\left(\frac{dV}{dH_{\mathrm{DC}}}\right) 
	= \frac{1}{d}\left(\frac{V_{\mathrm{OUT}}}{H_{\mathrm{AC}}}\right),
\end{equation}
where $d$ is the sample thickness. This approach enables determination of both linear and higher-order ME coefficients through nonlinear analysis.

Among multiferroic systems, strain-mediated composite materials exhibit significantly enhanced ME responses compared to single phase compounds~\cite{pradhan2020magnetoelectric}. In such systems, magnetostrictive strain generated in the magnetic phase under an applied magnetic field is elastically transferred across the interface to the piezoelectric phase, inducing electric polarization. The efficiency of this coupling depends on magnetic anisotropy, piezomagnetic sensitivity, piezoelectric strength, elastic compatibility, and interfacial strain transfer~\cite{petrov2007magnetoelectric}. In bilayer thin films and laminated composites, magnetoelectric coupling is strengthened by efficient interfacial strain transfer and elastic connectivity~\cite{kar2025giant}.

The relaxor ferroelectric 0.65Pb(Mg$_{1/3}$Nb$_{2/3}$)O$_3$-0.35PbTiO$_3$ (PMN-PT) exhibits exceptionally high piezoelectric coefficients and undergoes a ferroelectric-to-paraelectric phase transition near $T_{\mathrm{C}} \approx 450$~K, accompanied by a structural transformation (from tetragonal to cubic)~\cite{chen2014high}. NiFe$_2$O$_4$ (NFO), a ferrimagnetic inverse spinel with a high magnetic ordering temperature ($T_{\mathrm{N}} \sim$ 830~K), exhibits moderate magnetostriction and relatively low magnetic anisotropy~\cite{satish2023effect}, facilitating efficient strain generation at low magnetic fields~\cite{kaur2025introduction}. Owing to the combination of strong piezoelectricity in PMN-PT and favorable magnetic properties of NFO, the PMN-PT/NFO system serves as an attractive platform for investigating strain-mediated ME coupling~\cite{ahlawat2013evidence,ahlawat2017electric}. In contrast, the CoFe$_2$O$_4$ (CFO)-BaTiO$_3$ (BTO) composite represents a classical and extensively studied ME system~\cite{duong2007effect}. CFO exhibits large magnetostriction but possesses higher magnetocrystalline anisotropy~\cite{satish2023effect}, influencing its low-field magnetization dynamics and strain response, while BTO exhibits comparatively lower piezoelectric coefficients than PMN-PT~\cite{sebald2005morphotropic}. The contrasting magnetic softness and piezoelectric strength of the PMN-PT/NFO and CFO/BTO systems provide an excellent opportunity to systematically examine the material-dependent factors governing the magnitude and field dependence of strain-mediated magnetoelectric coupling.

Here, we report the development of a dynamic magnetoelectric measurement system, including a switch system for bipolar DC magnetic field capability, based on phase-sensitive lock-in detection and present a comparative investigation of strain-mediated ME coupling in PMN-PT/NFO and CFO/BTO composites. We would like to emphasize about a $180^{\circ}$ phase reversal upon field switching utilizing a home developed bipolar switching system. By correlating magnetic properties, piezoelectric characteristics, and dependency on DC and AC magnetic field of induced magnetoelectric voltage, this study provides deeper insight into the interplay between magnetic anisotropy, piezomagnetic response, and interfacial strain transfer in determining overall magnetoelectric performance.

\begin{figure}
	\centering
	\includegraphics[width=1\linewidth]{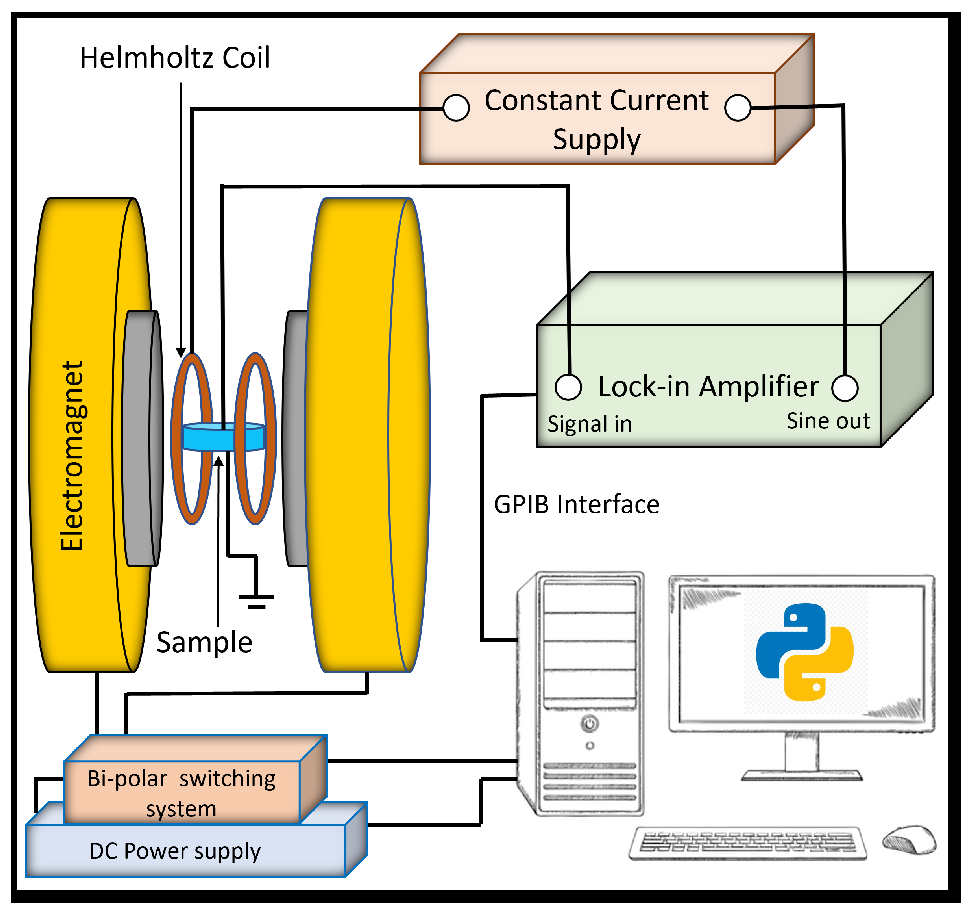}
	\caption{Schematics of the developed setup for magnetoelectric coupling measurement.}
	\label{SD-BI}
\end{figure}

\section{EXPERIMENTAL SECTION}
ME coupling measurements were performed using the dynamic lock-in technique~\cite{duong2007lock}. A schematic diagram of the experimental setup is shown in Fig.~\ref{SD-BI}. The AC magnetic field was generated using a Helmholtz coil consisting of 80 turns in each coil with a separation of 44~mm and a measured total resistance of 1.3~$\Omega$. The sample was mounted on a non-magnetic glass substrate and positioned at the center of the Helmholtz coil to ensure a uniform AC magnetic field distribution on the sample. The Helmholtz coil assembly was placed between the poles of a water-cooled DC electromagnet capable of producing magnetic fields up to 2~T when operated with an appropriate high-current power supply. In the present configuration, however, the electromagnet was driven using a Keithley 2260B unipolar power supply, limiting the accessible magnetic field range to approximately 6~kOe.
	
For bipolar magnetic field operation, a fully programmable, home-developed Arduino-based switching system was implemented to achieve precise and reproducible reversal of the DC magnetic field polarity up to $\pm$5.5~kOe. The circuit employed a mechanically switched relay-bridge configuration, integrated with flyback (backflow) diodes to suppress inductive voltage spikes during polarity transitions. Additionally, digital voltage and current indicators were incorporated for real-time monitoring, ensuring operational safety, accurate magnetic field control, and enhanced reliability during magnetoelectric measurements. All magnetoelectric measurements reported in this work were performed within a DC magnetic field range of $\pm$5~kOe.

A lock-in amplifier (Stanford Research Systems, Model SR830) was used both to provide the AC excitation signal and to detect the induced ME voltage. The internal sine-wave output of the lock-in amplifier (frequency range 1~mHz~-~102~kHz; amplitude range 0.004--5~V) was connected to the Helmholtz coil through a custom built voltage to constant current converter. This ensured a linear relationship between applied voltage and coil current (1~V $\equiv$ 1~A $\equiv$ 10~Oe), allowing generation of AC magnetic fields up to 40~Oe. The induced magnetoelectric voltage ($V_{\mathrm{OUT}}$) across the sample was measured using the same lock-in amplifier with appropriate sensitivity and time constant settings depending on the signal amplitude and measurement frequency. The entire system, including magnetic field control and data acquisition, was fully automated using Python scripts.

Polycrystalline 0.5(NFO)–0.5(PMN–PT) ceramics were fabricated by mixing the parent nanopowders of PMN–PT and NFO in an equimolar (1:1) ratio~\cite{ahlawat2013evidence}. The homogeneous mixture was ball-milled for 12~h to ensure uniform dispersion, followed by sintering at 750~$^\circ$C for 10~h. The nanocrystalline parent powders were synthesized separately via a sol–gel route, as reported elsewhere~\cite{ahlawat2011mossbauer,ahlawat2017electric}.
Another composit CFO was prepared by a conventional solid-state reaction method using CoO and Fe$_2$O$_3$ as starting precursors. The mixed powders were calcined at 900~$^\circ$C for 2~h and subsequently sintered at 1100~$^\circ$C to achieve phase formation and densification. The 0.5(CFO)–0.5(BTO) composite was obtained by thoroughly mixing the synthesized CFO powder with commercial BTO powder, followed by sintering at 1150~$^\circ$C for 4~h to form the final composite ceramic.

The synchrotron X-ray diffraction (SXRD) measurements were performed at the BL-13 beamline of the Indus-II synchrotron source using X-rays of wavelength $\lambda = 1.2398$~\AA{}. Diffraction data were collected with a Pilatus 200K 2D detector and converted to 1D patterns by radial integration for further analysis. Field emission scanning electron microscopy (FE-SEM) images ware captured to study morphology using FEI Nova NanoSEM 450 FE-SEM. Magnetic measurements were performed at room tempetrature by using a 14 T Vibrating sample magnetometer (VSM; Quantum Design Inc., USA). Prior to ME measurements, the samples were electrically poled along the thickness (thicknesses of the PMN-PT/NFO and CFO/BTO composites are 0.5 and 1.0 mm, respectively) direction at room temperature under a DC electric field of 2~kV\,mm$^{-1}$ for 1~h and subsequently short-circuited for 15~min to eliminate trapped charges.

\begin{figure}
	\centering
	\includegraphics[width=1\linewidth]{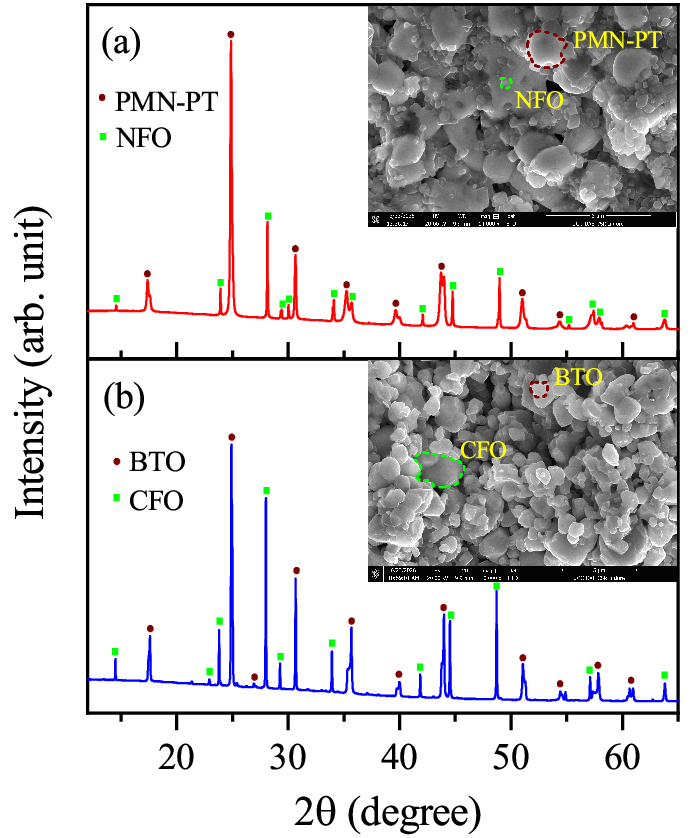}
	\caption{Synchrotron X-ray diffraction patterns of (a) PMN-PT/NFO and (b) CFO/BTO composites. The insets show the corresponding FE-SEM micrographs of the respective composites highlighting the distinct morphologies of the ferroelectric and ferrite phases.}
	\label{XRD}
\end{figure}

\begin{figure}
	\centering
	\includegraphics[width=1\linewidth]{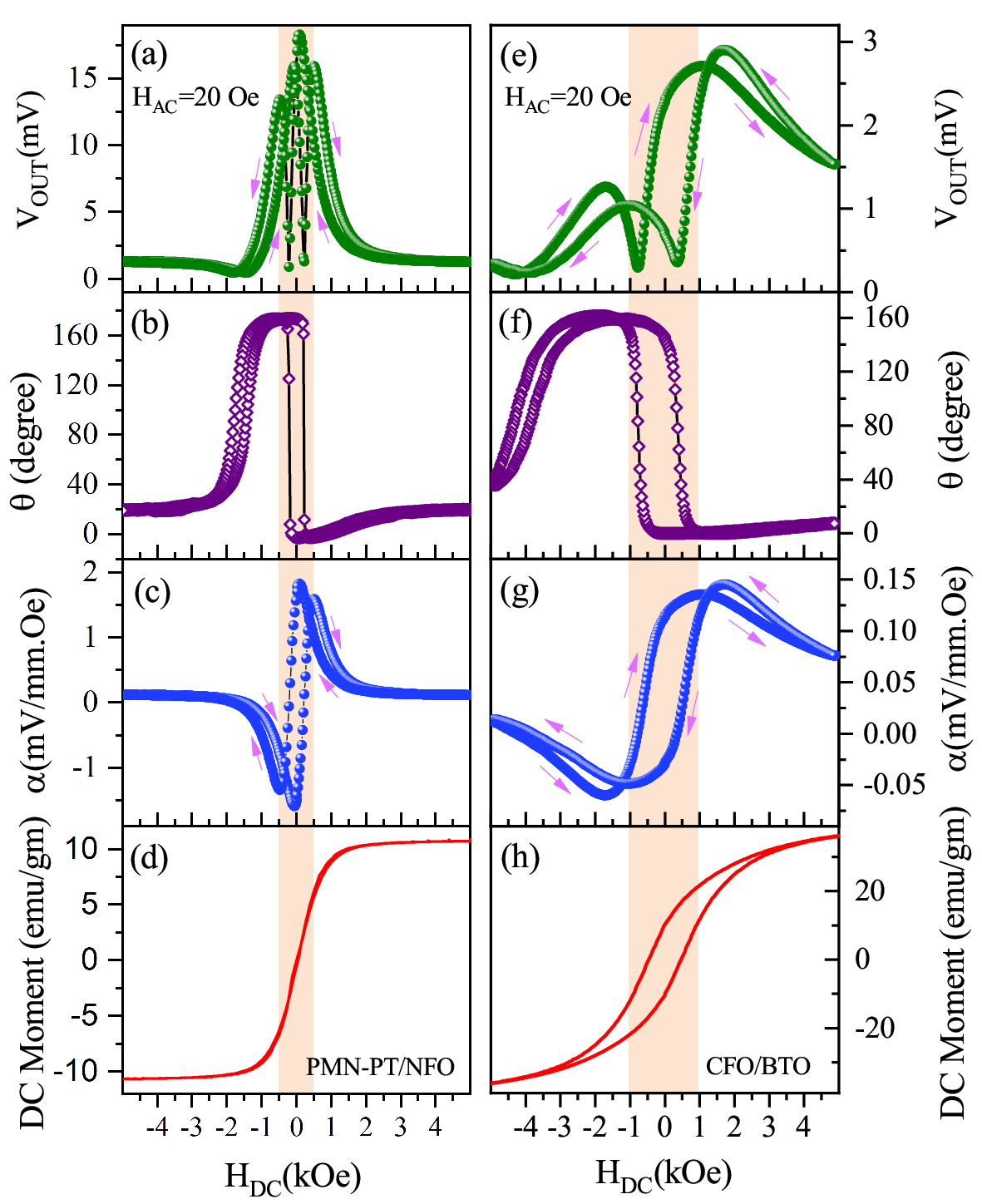}
	\caption{Magnetoelectric response and magnetic properties of (a-d) PMN-PT/NFO and (e-h) CFO/BTO showing (a,e) induced ME output voltage $V_{\mathrm{OUT}}$ as a function of applied DC magnetic field, (b,f) lock-in amplifier phase variation with DC magnetic field, (c,g) ME coupling coefficient $\alpha_{\mathrm{ME}}$ as a function of DC magnetic field up to 5~kOe measured at a fixed frequency of 23~Hz with an AC magnetic field amplitude of 20~Oe, and (d,h) room-temperature magnetic hysteresis ($M$-$H$) loops illustrating the correlation with ME response.}
	\label{com-1}
\end{figure}

\begin{figure}
	\centering
	\includegraphics[width=1\linewidth]{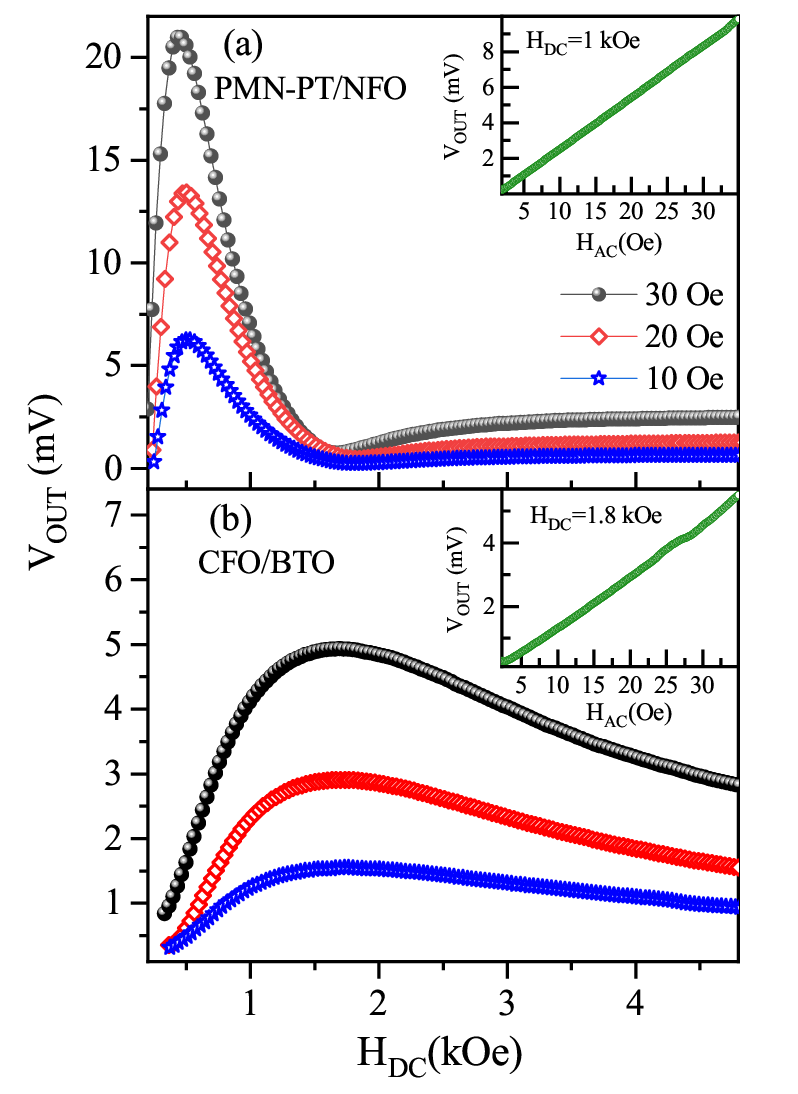}
	\caption{Magnetoelectric output voltage $V_{\mathrm{OUT}}$ as a function of DC magnetic field for (a) PMN--PT/NFO and (b) CFO/BTO, measured at a fixed frequency of 23~Hz with different AC magnetic field amplitudes of 10~Oe, 20~Oe, and 30~Oe; insets show the variation of ME output voltage $V_{\mathrm{OUT}}$ as a function of applied AC magnetic field in respective samples.}
	\label{com_me}
\end{figure}

\section{RESULTS AND DISCUSSION}

The room-temperature SXRD patterns of the PMN-PT/NFO and CFO/BTO composites are shown in Fig.~\ref{XRD}(a) and Fig.~\ref{XRD}(b), respectively. The diffraction data confirm the coexistence of ferroelectric (PMN-PT or BTO) and ferrimagnetic (NFO or CFO) phases without detectable secondary phases, consistent with earlier reports~\cite{ahlawat2017electric,nie2009effect}. All reflections are indexed to the perovskite and cubic spinel structures, confirming the successful formation of biphasic composites. The insets show the corresponding FE-SEM micrographs of the composites, where the ferrite and ferroelectric phases can be distinguished by their different grain-size distributions. Noticeably, in both systems, the ferrite phase exhibits sharper diffraction peaks than the ferroelectric phase, suggesting comparatively better crystallinity and lower microstrain in the spinel structure. In contrast, the broader PMN-PT and BTO reflections can be attributed to intrinsic lattice distortions and domain-related strain characteristic of ferroelectric perovskites. These microstructural features are expected to influence the efficiency of strain transfer across the ferrite-ferroelectric interface and, consequently, the strain-mediated magnetoelectric coupling. Lower microstrain and improved crystallinity in the ferrite phase facilitate the generation and transmission of magnetostrictive strain, whereas lattice distortion, structural heterogeneity, and elastic compatibility of the constituent phases govern the effectiveness of interfacial strain coupling and the resulting magnetoelectric response.

Equation (5) assumes that the applied AC magnetic field and the measured signal are perfectly in phase. In practice, however, the inherent L--R nature of the measurement circuit introduces a phase difference that increases with frequency, affecting the accurate evaluation of the ME coefficient~\cite{kuila2017measurement}. To minimize inductive effects and associated phase errors, the ME measurements were therefore performed at a low frequency (23~Hz).
At room temperature, the induced ME output voltage ($V_{\mathrm{OUT}}$) of the PMN-PT/NFO composite was measured as a function of DC magnetic field ($H_{\mathrm{DC}}$) up to $\pm 5$~kOe at a fixed frequency of 23~Hz with an AC magnetic field amplitude of 20~Oe, as shown in Fig.~\ref{com-1}(a). With increasing positive $H_{\mathrm{DC}}$, $V_{\mathrm{OUT}}$ initially increases and reaches a maximum value of 16~mV at 490~Oe, followed by a gradual decrease and saturation above $\sim$1.5~kOe. During the decreasing field cycle, a clear hysteresis is observed, with a higher maximum of 18.5~mV at approximately 70~Oe. A similar hysteretic behavior is obtained under negative magnetic fields, where the maximum $V_{\mathrm{OUT}}$ reaches 15.8~mV during the decreasing negative field sweep. The lock-in amplifier phase exhibits an abrupt $\sim180^\circ$ reversal upon switching the magnetic field polarity, confirming a sign change in the ME coupling (Fig.~\ref{com-1}(b)). The corresponding ME coupling coefficient ($\alpha_{\mathrm{ME}}$) extracted from in-phase component, displays an antisymmetric hysteretic behavior with respect to $H_{\mathrm{DC}}$. The maximum $\alpha_{\mathrm{ME}}$ values are $1.599$~mV/mmOe at 490~Oe during the increasing positive field cycle and $1.85$~mV/mmOe at $\sim$70~Oe during the decreasing cycle, while under negative fields the maximum magnitude reaches -1.58~mV/mmOe at -75~Oe (Fig.~\ref{com-1}(c)). The room-temperature magnetization ($M$--$H$) loop shows rapid magnetization at low fields and saturation above $\sim$1.5~kOe with a very small coercive field as shown in Fig.~\ref{com-1}(d), confirming the soft ferromagnetic nature of the composite. The peak in $\alpha_{\mathrm{ME}}$ coincides with the region of maximum $\frac{dM}{dH}$, indicating that the ME response is governed directly by the field induced variation of magnetization.

Under identical measurement conditions, the CFO/BTO composite exhibits a comparatively smaller ME response. Fig.~\ref{com-1}(e) shows that the $V_{\mathrm{OUT}}$ increases with increasing positive $H_{\mathrm{DC}}$, reaching a maximum of 2.725~mV at 1.06~kOe. During the decreasing cycle, a higher maximum of 2.929~mV is observed at $\sim$1.672~kOe, confirming hysteretic behavior. Under negative magnetic fields, the magnitude of $V_{\mathrm{OUT}}$ is comparatively reduced. The ME coefficient $\alpha_{\mathrm{ME}}$ also exhibits antisymmetric hysteresis, with a maximum value of 0.147~mV/mmOe during the decreasing positive field cycle, whereas under negative fields the magnitude reaches $-0.062$~mV/mmOe (see Fig.~\ref{com-1}(g)). The corresponding $M$-$H$ loop reveals gradual magnetization and saturation above $\sim$4.5~kOe, with a relatively large coercive field of $\sim$470~Oe (Fig.~\ref{com-1}(h)), indicating stronger magnetocrystalline anisotropy compared to NFO~\cite{satish2023effect}.

Fig.~\ref{com_me}(a,b) shows the dependence of $V_{\mathrm{OUT}}$ on DC magnetic field for different $H_{\mathrm{AC}}$ field values ($H_{\mathrm{AC}}$ = 10, 20, and 30~Oe) which further confirms the strain-mediated origin of the ME effect. For both composites, $V_{\mathrm{OUT}}$ increases linearly with increasing $H_{\mathrm{AC}}$, demonstrating operation within the linear magnetoelastic regime as shown in both insets of Fig.~\ref{com_me}.

In strain-mediated magnetic/ferroelectric composites, ME coupling arises from elastic interaction between the magnetostrictive and piezoelectric phases~\cite{nan1994magnetoelectric,livesey2011strain}. When a magnetic field is applied, magnetostrictive strain generated in the ferrite phase is mechanically transferred to the piezoelectric phase, producing an induced voltage. Consequently, the ME coefficient is proportional to the piezomagnetic coefficient $q = \frac{d\lambda}{dH}$ ($\lambda$ = magnetostriction coefficient) and the piezoelectric coefficient $d$, i.e., $\alpha_{\mathrm{ME}} \propto (q*d)$~\cite{pradhan2018exploring,bhoi2021unravelling}. The ME coefficient attains its maximum near the coercive field region, where domain wall motion and magnetization rotation are most active, and decreases at higher fields due to magnetic saturation. Reversal of magnetic field polarity changes the sign of magnetostrictive strain and results in an approximately $180^\circ$ phase shift in lock-in detection (Fig.~\ref{com-1}(c,f)), giving rise to an antisymmetric $\alpha_{\mathrm{ME}}$-$H_{\mathrm{DC}}$ response.

The significantly stronger ME response observed in PMN-PT/NFO compared to CFO/BTO as shown in Fig.~\ref{com-1}(b) can be attributed to the combined effect of magnetic softness and piezoelectric strength. NFO, being magnetically softer than CFO, exhibits lower anisotropy energy and larger effective $\frac{d\lambda}{dH}$ at low magnetic fields. Simultaneously, PMN-PT possesses a substantially higher piezoelectric coefficient than BTO~\cite{sebald2005morphotropic}. The synergistic combination of enhanced piezomagnetic sensitivity and strong piezoelectric response therefore leads to superior strain transfer efficiency and markedly enhanced ME coupling in PMN-PT/NFO. Overall, the strong low-field ME response, clear antisymmetric behavior, and linear AC field dependence confirm efficient strain-mediated coupling in both composites, with PMN-PT/NFO exhibiting significantly superior magnetoelectric performance.

\section{CONCLUSIONS}
In conclusion, a dynamic lock-in based magnetoelectric (ME) measurement system incorporating a home-developed switching unit for bipolar DC magnetic fields up to 5~kOe was successfully established to investigate strain-mediated coupling in composite multiferroics. Both PMN-PT/NFO and CFO/BTO exhibit ME coupling with clear antisymmetric behavior and a characteristic $180^{\circ}$ phase reversal upon magnetic field switching. The maximum ME coefficient of PMN-PT/NFO ($\alpha_{\mathrm{ME}} \approx 1.85$~mV\,mm$^{-1}$\,Oe$^{-1}$) is significantly higher than that of CFO/BTO ($\approx 0.147$~mV\,mm$^{-1}$\,Oe$^{-1}$), demonstrating superior low-field performance. This enhancement originates from the magnetic softness of NFO combined with the high piezoelectric response of PMN-PT, which together promote efficient strain transfer. In contrast, the relatively weaker response of CFO/BTO is constrained by higher magnetic anisotropy and lower piezoelectric sensitivity. Overall, these findings highlight the critical role of simultaneously optimizing piezomagnetic and piezoelectric properties to achieve enhanced magnetoelectric performance in composite systems.

\bibliography{references}
\bibliographystyle{aipnum4-2}

\end{document}